\documentclass[10pt,showpacs,floatfix,letterpaper,amsmath,amsfonts,amssymb,aps,pra,reprint,superscriptaddress]{revtex4-1}
\usepackage[utf8]{inputenc}
\usepackage{hyperref}
\usepackage{ulem}
\usepackage{times}
\usepackage{amssymb}
\usepackage{amsmath}
\usepackage{epsfig}
\usepackage{wasysym}
\usepackage{color}
\usepackage{pifont}
\usepackage{colortbl}
\usepackage{color}
\usepackage{epsfig}
\usepackage{tabularx}
\usepackage{graphicx}
\usepackage{amsfonts}
\usepackage{euscript}
\usepackage{verbatim}
\usepackage{latexsym}
\usepackage[T1]{fontenc}
\usepackage{mathtools}
\usepackage{float}

\newcommand\bea{\begin{eqnarray}}
\newcommand\eea{\end{eqnarray}}
\newcommand\beq{\begin{equation}}
\newcommand\eeq{\end{equation}}
\newcommand\bsq{\begin{subequations}}
\newcommand\esq{\end{subequations}}

\begin{document}
\title{Quantum error correction using weak measurements}
\author{Parveen Kumar}
\email {parveenkumar@chep.iisc.ernet.in}
\author{Apoorva Patel}
\email {adpatel@chep.iisc.ernet.in}
\affiliation{Centre for High Energy Physics, Indian Institute of Science,
Bangalore 560012, India}
\date{\today}
\begin{abstract}
The standard quantum error correction protocols use projective measurements
to extract the error syndromes from the encoded states. We consider the more
general scenario of weak measurements, where only partial information about
the error syndrome can be extracted from the encoded state. We construct a
feedback protocol that probabilistically corrects the error based on the
extracted information. Using numerical simulations of one-qubit error correction
codes, we show that our error correction succeeds for a range of the weak
measurement strength, where (a) the error rate is below the threshold beyond
which multiple errors dominate, and (b) the error rate is less than the rate
at which weak measurement extracts information. It is also obvious that error
correction with too small a measurement strength should be avoided.
\end{abstract}

\maketitle

\section{Introduction}

In recent years, the field of quantum information and quantum computation has
rapidly progressed from a theoretical framework to an experimental level,
where toy systems carry out simple but practical tasks. The main hurdle to
be overcome for large scale integration of quantum devices is a control over
errors. No physical system can be perfectly isolated from the environment,
and the inevitable disturbances affect its operation. Quantum information
processors are especially sensitive in this regard, and designs that would
make them fault-tolerant are an outstanding challenge. A recent road map for
fault-tolerant quantum computation \cite{Devoret1169} emphasizes the role that
quantum error correction (QEC) would have to play to protect the quantum data.
The QEC strategy is to redundantly encode the quantum information in a larger
Hilbert space, such that the logical qubits experience a significantly smaller
error rate than what the physical qubits do. A cascade of QEC codes can then
make the lifetime of encoded quantum information as long as desired.

The standard QEC codes are illustrated by the label $[\![n,k,d]\!]$. They
encode $k$ logical qubits into $n$ physical qubits, and $d$ is the minimum
distance between logical codewords. The errors are discretized to a finite
set in the Pauli operator basis for each qubit, they are detected using
projective measurements of the appropriate syndromes, and then the
measurement-result-dependent inverse transformations restore the original
information. This procedure corrects upto $[{d-1\over2}]$ Pauli errors,
and the residual error rate of the encoded state is given by the probability
of having more than $[{d-1\over2}]$ errors. The procedure is worthwhile only
when the error rate of $n$-qubit encoded state is smaller than the error
rate of $k$-qubit unencoded state, and that happens only when the error
rate of unencoded state is below a critical threshold. Such codes were
first devised by Shor \cite{ShorPhysRevA.52.R2493} and Steane
\cite{SteanePhysRevLett.77.793}, and a variety of them have been constructed
since then. For practical applications, it is paramount to understand the
error mechanisms as well as possible, and then design the codes to maximize
the critical threshold. Attempts to build quantum error correction procedures
for several physical systems have been made, e.g. liquid
\cite{CoryPhysRevLett.81.2152,KnillPhysRevLett.86.5811,BoulantPhysRevLett.94.130501}
and solid state \cite{MoussaPhysRevLett.107.160501} NMR, trapped ions
\cite{chiaverini2004realization,Schindler1059}, photon modes
\cite{PittmanPhysRevA.71.052332}, superconducting qubits
\cite{reed2012realization,kelly2015state}, and NV centers in diamond
\cite{waldherr2014quantum,taminiau2014universal}. 

Projective measurements are instantaneous, they extract maximum information
about the measured observable, and their post-measurement state is known with
certainty. These properties allow accurate error correction. In contrast,
weak measurements are performed on a stretched out time scale, with only a
gentle disturbance to the quantum system \cite{AharonovPhysRevLett.60.1351}.
They extract only partial information about the measured observable, which
rules out complete error correction. With weak measurements, therefore, we
can only aim to restore the quantum state with as high fidelity as possible.
In this work, we present a protocol to implement quantum error correction
using weak measurements. Obviously, it would be useful only when projective
measurements cannot be carried out for some reason.

Attempts to construct QEC protocols using weak measurements have been made
before \cite{AhnPhysRevA.65.042301,SarovarPhysRevA.69.052324}. In our work,
we use continuous stochastic measurement dynamics to design a QEC feedback
protocol. We propose a general feedback scheme based on binary weak
measurements, and numerically investigate its efficacy as a function of
the measurement coupling. Our protocol is appropriate for weak measurements
of superconducting transmon qubits, but it can be easily extended to other
physical systems.

This paper is organized as follows. Section II briefly reviews how a quantum
system evolves during weak measurement, using the setting of circuit QED, and
presents our feedback scheme for a quantum register when all measurements are
binary weak measurements. Section III describes the numerical simulation
results of our protocol, for the bit-flip error correction of a single qubit
encoded in a three-qubit register, and arbitrary error correction of a single
qubit encoded in a five-qubit register. We conclude with a discussion of our
results in Section IV.

\section{Weak measurements and feedback}

A quantum system interacting with the environment and the measuring apparatus
undergoes a complex evolution. We omit any driving term, e.g. the system could
be some quantum information stored in memory, and model the total evolution as:
\begin{equation}
  \label{1qee3}
  \frac{d}{dt} \rho = i[\rho,H_E+H_F] + M[P_i,\rho].
\end{equation}
Here $H_E$ and $H_F$ are the error and the feedback Hamiltonians respectively,
and $P_i$ are the projectors for the measured observable. Compared to the
usual framework for QEC codes \cite{Nielsen}, we have replaced the projective
measurement evolution, $\rho\rightarrow\sum_i P_i\rho P_i$, with the weak
measurement evolution operator $M[P_i,\rho]$.

We use the framework of continuous quantum stochastic dynamics to describe
weak measurements \cite{GisinPhysRevLett.52.1657}. In this framework, an
ensemble of quantum trajectories is generated, by combining geodesic evolution
of the initial quantum state to the eigenstates $|i\rangle$ of the measured
observable with white noise fluctuations. In this evolution, every quantum
trajectory keeps a pure state pure (i.e. preserves $\rho^2=\rho$), and the
Born rule is satisfied at every instant of time (upon averaging over the
stochastic noise). We use the notation \cite{patel2015weak}:
\begin{equation}
\label{ce2}
M[P_i,\rho] = g \sum_i w_i [\rho P_i + P_i \rho - 2\rho\;\textrm{tr}(P_i \rho)],
\end{equation}
where the system-apparatus interaction parameter $g$ has dimensions of energy
($g$ can be time-dependent, in which case $g\tau$ in the rest of the article
should be interpreted as $\int_0^{\tau}g\;dt$). $M[P_i,\rho]$ vanishes at the
fixed points $\rho_i^* = P_i$, ensuring termination and repeatability of
measurements. The weights $w_i$ are normalized to $\sum_i w_i=1$. They are
chosen such that the system's dynamics reproduces the well-established quantum
behaviour, and the weak measurement contributes a stochastic noise to $w_i$
\cite{KorotkovPhysRevB.60.5737,KorotkovPhysRevB.63.115403,vijay2012stabilizing,murch2013observing}.
The projective measurement is recovered in the limit $g\tau\rightarrow\infty$.

\subsection{Binary measurement}

For a binary weak measurement,
\begin{equation}
\label{mc}
w_0 - w_1 = \textrm{Tr}(\rho P_0) - \textrm{Tr}(\rho P_1) + \frac{1}{\sqrt{g}} \xi(t),
\end{equation}
where $\xi (t)$ is a white noise with $\langle\xi(t)\rangle = 0$ and
$\langle\xi(t)\xi(t')\rangle = \delta(t-t')$. During weak measurement,
$w_0-w_1$ can be experimentally observed along any quantum trajectory.

In recent years, weak measurements have been implemented experimentally for
superconducting transmon qubits \cite{vijay2012stabilizing,murch2013observing}.
A transmon qubit is essentially a tunable nonlinear quantum oscillator made
of two Josephson junctions in a superconducting loop shunted by a capacitor.
The lowest two energy levels of the nonlinear oscillator are used as a qubit.
The qubit is kept in a microwave cavity with dispersive coupling, and its weak
measurements are carried out by probing the cavity by a microwave signal.
For weak measurement of a transmon, the signal observed by the apparatus is
a current $I_m$, and $w_0-w_1$ is obtained by scaling it suitably. We adopt
the convention that the ideal measurement current is $\pm\frac{\Delta I}{2}$
for the measurement eigenstates $P_0$ and $P_1$. Then the scaled measurement
current, $\frac{2I_m}{\Delta I}=w_0-w_1$, provides an estimate of
$\textrm{Tr}(\rho P_0) - \textrm{Tr}(\rho P_1)$.

\begin{figure}
\centering
\includegraphics[width=8.5cm]{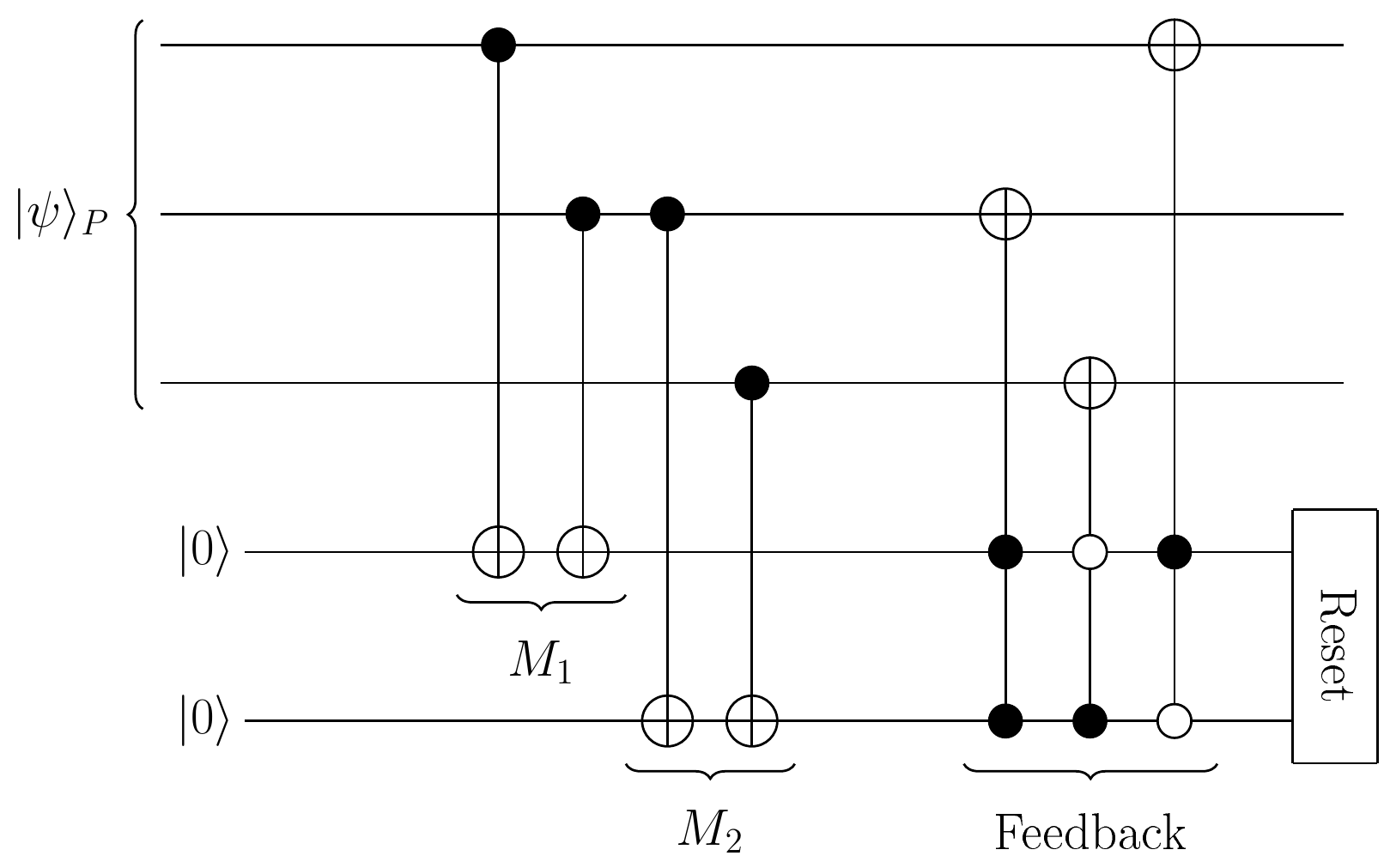}
\caption{Quantum logic circuit to correct a single bit-flip error,
for a logical qubit encoded in three physical qubits. Two ancilla qubits,
initialised in the ground state, are coupled to the three physical qubits,
and two parities of qubit-pairs are extracted. These parities determine the
location of the error without disturbing the encoded logical state. Using
the syndrome information, the error is corrected by applying controlled
inverse transformations. Finally, the ancilla qubits are reset, which
disentangles them from the physical qubits.}
\label{bit_flip_QEC_circuit}
\end{figure}

The redundancy of the encoding QEC protocol allows extraction of the syndrome
information from the encoded state without disturbing the encoded information.
The simplest code to correct a single bit-flip error encodes the logical qubit
into three physical qubits. Parities of qubit-pairs provide the syndrome
information, which is extracted into two additional ancilla qubits, using
$C$-NOT logic gates as illustrated in Fig.\ref{bit_flip_QEC_circuit}.
These parities are then used to apply controlled inverse logic operations
to eliminate the error. (For the sake of simplicity, we have assumed that
all gate operations are perfect, and the qubits are exposed to the error
perturbations only before the measurement and feedback steps. A more
elaborate fault-tolerant prescription can take care of errors occurring
anywhere during the evolution.) After using the parities for the feedback,
the ancilla qubits need to be disentangled from the encoded state, and
that is achieved by resetting the ancilla qubits. This resetting is an
irreversible step, and it reduces the fidelity of the encoded quantum state.
Different methods have been proposed for resetting the ancilla qubits
\cite{doi:10.1063/1.3435463,GeerlingsPhysRevLett.110.120501}, and they
can be applied equally well after strong or weak measurements.

\subsection{Feedback design}

Initially the quantum system is in the logical subspace. Under the influence
of undesired disturbances, it moves out of the logical subspace along some
of the error directions. The syndrome operators are designed to reveal the
information about the error without disturbing the encoded signal. The binary
syndrome operators have a $+1$ eigenvalue for the logical subspace and a $-1$
eigenvalue for some of the error states. The complete error information is
constructed from a collection of binary syndrome measurements.

The measurement evolution, described by Eqs. (\ref{1qee3},\ref{ce2},\ref{mc}),
can be expressed in the It\^o form as \cite{GisinPhysRevLett.52.1657}:
\begin{eqnarray}
\label{dz}
dz &=& g\;\tanh z\;dt + \sqrt{g}\;dW(t)\;,\\ 
\tanh z &=& \textrm{Tr}(\rho P_0) - \textrm{Tr}(\rho P_1) \equiv 2p-1 .
\end{eqnarray}
Here the stochastic Wiener increment obeys $\langle dW(t)\rangle = 0$ and
$\langle (dW(t))^2 \rangle = dt$. For any initial state $z_0$, the time
evolution of the quantum trajectory distribution is known
\cite{GisinPhysRevLett.52.1657}.
After time $\tau$, the initial $\delta$-function distribution becomes a sum
of two Gaussians with areas $p_0$ and $1-p_0$, centered at $z_0+g\tau$ and
$z_0-g\tau$ respectively, and with a common variance $g\tau$.

The unperturbed logical space state has $p=1$ or $z=\infty$. Due to errors,
the state gets shifted to some finite $z_0$, and the feedback task is to
move it back to $z=\infty$. The syndrome measurement is designed to estimate
$z_0$, but ends up shifting the state more in the process. To figure out what
feedback to apply, with only incomplete information available, we make the
following approximations:\\
(i) We approximate the measurement evolution as a binary walk on a line,
producing $\delta$-function distributions at $z_0\pm g\tau$ with probabilities
$p_0$ and $1-p_0$. For each step of a quantum trajectory, only one of the two
possibilities occur, i.e. the post-measurement state is either $z_0+g\tau$ or
$z_0-g\tau$. Of course, $g\tau$ is known from the properties of the measurement
apparatus.\\
(ii) The current measured by the apparatus provides an average value of the
signal over the measurement duration; so we use $\frac{2I_m}{\Delta I}$ to
estimate either $z_0 + \frac{g\tau}{2}$ or $z_0 - \frac{g\tau}{2}$.\\
(iii) The measured current $I_m$ is a combination of
$\textrm{Tr}(\rho\sigma_z)$ and white noise. We next assume that if
$I_m > 0$ then the post-measurement state is $z_0+g\tau$, and if $I_m < 0$
then the post-measurement state is $z_0-g\tau$.

With these assumptions, we convert $I_m$ into a rotation angle on the Bloch
sphere, and design the feedback transformation. When $I_m > 0$, the act of
measurement moves the state from $z_0$ to $z_0+g\tau$, which is closer to
the logical subspace corresponding to $z=\infty$. We then opt to apply no
additional feedback operation. When $I_m < 0$, we construct the feedback
operation as follows. In terms of polar coordinates on the Bloch sphere,
\begin{equation}
\textrm{Tr}(\rho P_0) = \frac{1+\cos\theta}{2},\;\;\textrm{Tr}(\rho P_1) = \frac{1-\cos\theta}{2}.
\end{equation}
We estimate the post-measurement state location as,
\begin{eqnarray}
\label{1qtanhz2}
\cos\theta &=& \tanh(z_0-g\tau) \nonumber \\
             &=& \frac{\tanh\left(z_0-\frac{g\tau}{2}\right) - \tanh\left(\frac{g\tau}{2}\right)}{1-\tanh\left(z_0-\frac{g\tau}{2}\right) \tanh\left(\frac{g\tau}{2}\right)} \nonumber \\
             &\rightarrow& \frac{\frac{2I_m}{\Delta I} - \tanh\left(\frac{g\tau}{2}\right)}{1-\frac{2I_m}{\Delta I} \tanh\left(\frac{g\tau}{2}\right)} \;.
\end{eqnarray}
Here the replacement of $\tanh(z_0-\frac{g\tau}{2})$ by $\frac{2I_m}{\Delta I}$
is the most severe approximation, due to which the last line of the previous
equation is not restricted to the interval $[-1,1]$. Whenever the value goes
outside this interval, we replace it by $-1,+1$ corresponding to the value
being less than $-1$ or greater than $+1$ respectively.

The rotation angle $\theta$ represents the combined shift of the quantum
state due to error and measurement. This shift can be reversed by applying
an inverse rotation by $-\theta$. The binary measurement does not determine
the polar angle $\phi$ of the shift on the Bloch sphere. To estimate $\phi$,
knowledge of off-diagonal elements of $\rho$ is needed, which is not available
to us. (In case of projective measurements, $\theta=0,\pi$ and ignorance of
$\phi$ does not matter.) Knowledge of $\phi$ is necessary to decide around
which axis in the X-Y plane the feedback rotation $-\theta$ should be applied.
Without that knowledge, we ad hoc choose the X-axis as the rotation axis.
Consequently, the feedback may correct the error or may make it worse.
Whenever the error gets worse, it would become easier to detect, and then
to correct, in the next iteration. We hope that the procedure converges with
repeated error correction steps with high fidelity, and our simulation results
support that.

In weak measurements, the measured current, given by Eq.(\ref{mc}), is a
noisy current. We can reduce the noise, and hence increase the accuracy
of Eq.(\ref{1qtanhz2}), by accumulating the signal over as long evolution
time as possible:
\begin{equation}
\label{fc}
\tilde{I}_m = \int_0^\tau I_m(t')\;dt'.
\end{equation}
Use of the integrated current $\tilde{I}_m$ as the signal, instead of the
original current $I_m$, is tantamount to using a stronger interaction
strength $g$. Also, a non-uniform integration weight in the definition of
$\tilde{I}_m$ would lose some information, and so is not worthwhile.
Once the feedback has been applied, then we have to erase all the current
history, and wait until ample new current data is accumulated, before
deciding on the next feedback operation. All this is equivalent to saying
that we apply feedback only when we have sufficient information, and don't
disturb the system otherwise.

\subsection{Multiple binary measurements}

Now consider the situation where the syndrome consists of two commuting binary
measurements, and $I_m^1$ and $I_m^2$ are the corresponding measured currents.
The physical Hilbert space can be divided into four sectors corresponding to
the measurement projectors $P_0^1 P_0^2$, $P_1^1 P_0^2$, $P_0^1 P_1^2$,
$P_1^1 P_1^2$, where the superscript denotes the measurement number. For each
binary measurement, we determine the rotation angle as in the previous
subsection:
\begin{eqnarray}
I_m^k > 0 &\Rightarrow& \theta = 0 \;, \nonumber \\
I_m^k < 0 &\Rightarrow& \cos\theta_k = \left[\frac{\frac{2I_m^k}{\Delta I} - \tanh\left(\frac{g\tau}{2}\right)}{1-\frac{2I_m^k}{\Delta I}\tanh\left(\frac{g\tau}{2}\right)}\right]_r,
\end{eqnarray}
where $[\ldots]_r$ denotes reduction of the value to the interval $[-1,1]$,
as described after Eq.(\ref{1qtanhz2}). The feedback operation is then
constructed from these angles. The syndrome definition tells us the location
of the error in the multi-qubit register; so we determine the location of the
error from the signs of $\{I_m^k\}$. For $\sigma_z$-measurements, we do not
have knowledge of the polar angles $\phi_k$, and as before, we ad hoc choose
the $X$-axis as the rotation axis.

When all the $I_m^k$ are positive, we do not apply any feedback operation.
When only one of the $\{I_m^k\}$ is negative, we use the inverse rotation
$-\theta_k$ as the feedback operation. When more than one $\{I_m^k\}$ are
negative, assuming each $I_m^k < 0$ to be an equal diagnostic of the error,
we estimate movement of the state out of the logical subspace by averaging
the corresponding $\cos\theta_k$. The feedback operation is then the inverse
rotation $-\bar{\theta}$ obtained from the averaged projection. For instance,
when $I_m^1 < 0$ and $I_m^2 < 0$, 
\begin{eqnarray}
\cos\bar{\theta} &=& \frac{1}{2}\left(\cos\theta_1 + \cos\theta_2\right) .
\end{eqnarray}
This is an empirical prescription, but numerically we find it to be a good
approximation.

\section{Numerical simulations and results}

We have performed numerical simulations to ascertain the accuracy of our
feedback scheme. Various terms on the r.h.s. of the evolution equation
(\ref{1qee3}) contribute simultaneously in reality. For ease of simulation,
however, we calculate these contributions one by one for evolution time $\tau$, and then combine them together. As per Trotter's formula, the error incurred in this procedure is $O(\tau^2)$, and we make that inconsequential by choosing $\tau$ to be sufficiently small. 

Within time step $\tau$, the system evolution is broken up into three parts:
(1) evolution under error, (2) evolution under measurement, and
(3) evolution under feedback. For evolution under error, we evolve
Eq.(\ref{1qee3}) using the fourth order Runge-Kutta method and only the
$H_E$ contribution.

We then include the effect of measurement by performing a probabilistic
Bayesian update \cite{korotkov2011quantum} that effectively integrates
Eq.(\ref{ce2}). The measurement current
$I_m = \frac{1}{\tau}\int^{\tau}_0 I(t^\prime)dt^\prime$ is drawn from
Gaussian probability distributions centered at $\pm\frac{\Delta I}{2}$
and with standard deviations $\sigma = \frac{\Delta I}{2\sqrt{g\tau}}$.
Explicitly, the conditional probability distributions for the measurement
current, when the system is in $|0\rangle$ or $|1\rangle$ state, are
\cite{vijay2012stabilizing}:
\begin{eqnarray}
\label{cpcd}
P(I_m||0\rangle) &=& \frac{1}{\sqrt{2\pi\sigma^2}}\;\exp{\left[-\frac{(I_m - \frac{\Delta I}{2})^2}{2\sigma^2}\right]}, \nonumber \\
P(I_m||1\rangle) &=& \frac{1}{\sqrt{2\pi\sigma^2}}\;\exp{\left[-\frac{(I_m + \frac{\Delta I}{2})^2}{2\sigma^2}\right]}.
\end{eqnarray}
With these conditional distributions, the probability distribution for the
measurement current is,
\begin{equation}
\label{pcd}
P(I_m) = \textrm{Tr}(\rho(\tau) P_0) P(I_m||0\rangle) + \textrm{Tr}(\rho(\tau) P_1) P(I_m||1\rangle).
\end{equation} 

For a binary measurement of an $N$-dimensional system, all $\rho_{ii}$ that
appear in $\textrm{Tr}(\rho P_0)$ are updated as
\cite{KorotkovPhysRevB.63.115403}:
\begin{eqnarray}
\label{deuf1}
\rho_{ii}(\tau) = \frac{\rho_{ii}(0)\;P(I_m||0\rangle)}{P(I_m)} ,
\end{eqnarray}
while all $\rho_{jj}$ that appear in $\textrm{Tr}(\rho P_1)$ are updated as:
\begin{eqnarray}
\label{deuf2}
\rho_{jj}(\tau) = \frac{\rho_{jj}(0)\;P(I_m||1\rangle)}{P(I_m)} .
\end{eqnarray}
All the off-diagonal $\rho_{ij}$ are updated according to the unitary
evolution constraint,
\begin{equation}
\rho_{ij}(\tau) = \rho_{ij}(0)\sqrt{\frac{\rho_{ii}(\tau) \; \rho_{jj}(\tau)}{\rho_{ii}(0) \; \rho_{jj}(0)}} .
\end{equation}

Finally, we apply the feedback transformation described in the previous
section, by converting the rotation into the feedback Hamiltonian $H_F$,
and evolving Eq.(\ref{1qee3}) with only the $H_F$ contribution.

\subsection{Single qubit bit-flip error correction}

Consider a single qubit system initialized in the state $|0\rangle$ that
undergoes a bit-flip error. The error Hamiltonian is
\begin{equation}
\label{1qeh}
H_E = \gamma \sigma_x~,
\end{equation}
where $\sigma_x$ is the Pauli operator representing the bit-flip error,
and the error coupling $\gamma$ is chosen as a Gaussian random number with
zero mean and variance $\alpha^2$. (We choose this form of $\gamma$ as
a likely scenario in actual experiments.) Evolution under the error
Hamiltonian for time $\tau$ changes the state of the system as
\begin{equation}
\label{re1q}
\rho(\tau) = U~\rho(0)~U^{\dagger}~,
\end{equation}
where $U=e^{-i \gamma \sigma_x \tau}$. For a system initialized in the state
$|0\rangle$, Eq.(\ref{re1q}) simplifies to
\begin{eqnarray}
\label{re1q1}
\rho(\tau) &=& 
        \begin{pmatrix}
        \cos^2 (\gamma\tau) & i \sin(\gamma\tau)\cos(\gamma\tau) \\
        -i \sin(\gamma\tau)\cos(\gamma\tau) & \sin^2 (\gamma\tau)
        \end{pmatrix} .
\end{eqnarray}
Averaging over the distribution of the error coupling $\gamma$, the averaged
density matrix becomes:
\begin{eqnarray}
\langle\rho(\tau)\rangle &=& \int_{-\infty}^{\infty}\frac{1}{\sqrt{2\pi\alpha^2}} e^{-\gamma^2/\alpha^2} ~ \rho ~ d\gamma ~, \nonumber \\
       &=&
       \label{re1q2}
       \frac{1}{2}
       \begin{pmatrix}
       1+e^{-2 (\alpha \tau)^2} & 0 \\
       0 & 1-e^{-2 (\alpha \tau)^2}
       \end{pmatrix}~.       
\end{eqnarray}
Fidelity of the quantum state evolves as
\begin{equation}
\label{fid_1q_bitflip_error}
f^g_{\textrm{err}} = \langle \cos^2(\gamma\tau) \rangle = \frac{1}{2}\left(1+e^{-2 (\alpha \tau)^2}\right) ~,
\end{equation}
and eventually the system reaches the completely mixed state.

\begin{figure}
\centering
\includegraphics[width=0.5\textwidth]{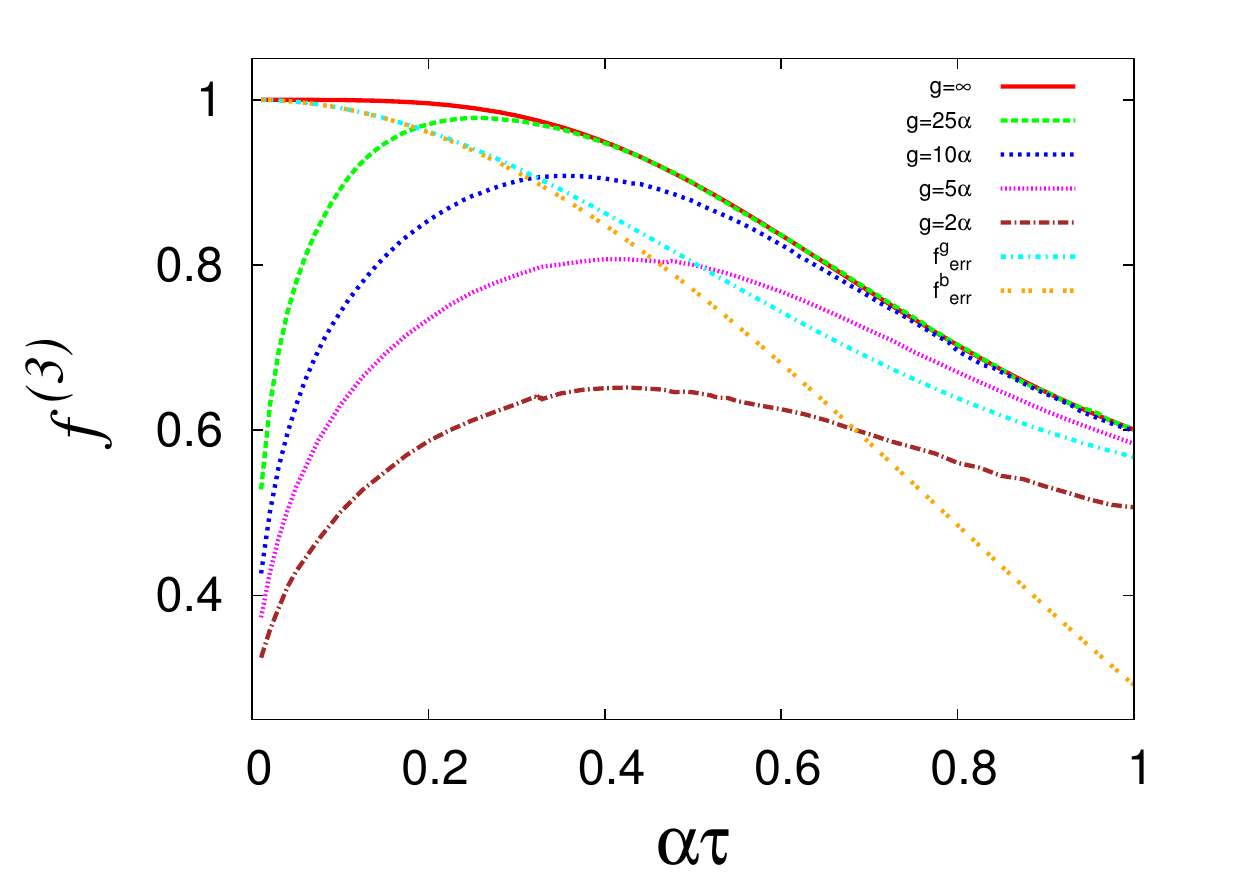}
\caption{Behaviour of the fidelity as a function of the error $\alpha\tau$
for the single qubit bit-flip error correction model. The orange and the cyan
curves show the fidelity in absence of any measurement or error correction,
for binary and Gaussian error distributions respectively. The red curve shows
the fidelity of the three-qubit register, when we perform error correction
using projective measurement. Rest of the curves show the fidelity for various
values of the measurement coupling. Clearly, the fidelity improves with
increasing measurement coupling, and approaches the one obtained by projective
measurement error correction as $g\tau\rightarrow\infty$.}
\label {fig_bfe_f_vs_setau}
\end{figure}

Choosing $\gamma$ as a Gaussian random variable is an assumption, and the
error distribution may be different in a different experimental setting.
As another example, we consider the binary error distribution case, again
with zero mean and variance $\alpha^2$. Then the error coupling is either
$+\alpha$ or $-\alpha$ with equal probability. In this case, the averaged
density matrix after time evolution $\tau$ becomes:
\begin{eqnarray}
\langle\rho(\tau)\rangle &=& 
       \label{re1q3}
       \begin{pmatrix}
       \cos^2 (\alpha\tau) & 0 \\
       0 & \sin^2 (\alpha\tau)
       \end{pmatrix}~,       
\end{eqnarray}
and the fidelity is
\begin{equation}
\label{fid_1q_bitflip_error_bm}
f^b_{\textrm{err}} = \cos^2 (\alpha\tau)~.
\end{equation}
Although we use the Gaussian error distribution in our simulations throughout
this work, we note that the choice between the Gaussian or the binary error
distribution does not matter much for small values of $\alpha\tau$, both
giving essentially the same fidelity as shown in Fig. \ref{fig_bfe_f_vs_setau}.

To protect a single qubit system against bit-flip error, we redundantly
encode it in a three-qubit register as:
\begin{eqnarray}
\label{3qenc}
|0\rangle_L &\rightarrow& |000\rangle_P ~, \nonumber \\
|1\rangle_L &\rightarrow& |111\rangle_P ~.
\end{eqnarray} 
The states $|0\rangle_L$ and $|1\rangle_L$ are basis vectors of the logical
space, while $|000\rangle_P$ and $|111\rangle_P$ are basis vectors of the
physical space. A general logical state to be protected is
\begin{equation}
\label{3qis}
|\psi\rangle = a |0\rangle_L + b |1\rangle_L,
\end{equation}
with $|a|^2 + |b|^2 = 1$. In our simulations, we have chosen the
initial state as $a = 1$ to simplify calculations; the linearity of
evolution ensures that when the protocol works for the basis states,
it will also work for any superposition of basis states. Our measured
fidelity of the encoded state is thus
\begin{equation}
\label{fid_enc_1q_bit_flip}
f^{(3)}={}_P\langle 000|\rho|000\rangle_P ~.
\end{equation} 
We diagnose the bit-flip errors using the syndrome operators $ZZI$
(i.e. $\sigma_z\otimes\sigma_z\otimes I$) and $IZZ$
(i.e. $I\otimes\sigma_z\otimes \sigma_z$). 

In the three-qubit Hilbert space, the independent bit-flip error Hamiltonian
is:
\begin{equation}
H_E = \gamma_1 (XII) + \gamma_2 (IXI) + \gamma_3 (IIX),
\end{equation}
where the error couplings $\gamma_i$ are independent Gaussian random numbers
with zero mean and variance $\alpha^2$.

Let the measurement currents for $ZZI$ and $IZZ$ be $I_m^1$ and $I_m^2$
respectively. We take the feedback Hamiltonian to be
\begin{equation}
H_F = \lambda_1 (XII) + \lambda_2 (IXI) + \lambda_3 (IIX),
\end{equation}
where $\lambda_i$ are the feedback couplings. We estimate the feedback
rotation angles as described in Section II.C. The feedback couplings are then:
\begin{eqnarray}
\label{3qfc}
\lambda_1 &= 
\begin{cases}
-\frac{\theta_1}{2\tau} & \mathrm{if} \; I_m^1 < 0 \;\mathrm{and}\; I_m^2 > 0, \nonumber\\
0 & \mathrm{otherwise},
\end{cases} \nonumber \\
~\nonumber \\ 
\lambda_2 &= 
\begin{cases}
-\frac{\bar{\theta}}{2\tau} & \mathrm{if} \; I_m^1 < 0 \;\mathrm{and}\; I_m^2 < 0, \\
0 & \mathrm{otherwise},
\end{cases} \\
~\nonumber \\ \nonumber
\lambda_3 &= 
\begin{cases}
-\frac{\theta_2}{2\tau} & \mathrm{if} \; I_m^1 > 0 \;\mathrm{and}\; I_m^2 < 0,  \nonumber \\
0 & \mathrm{otherwise}.
\end{cases} 
\end{eqnarray}

In case of projective measurement (i.e. $g\tau\rightarrow\infty$),
$\cos\theta_i = \pm 1$, and the fidelity of the quantum state after
applying the feedback can be analytically determined as:
\begin{equation}
\label{fid_bit_flip_proj_meas}
f^g_{\textrm{ec}} = \frac{1}{4}\left(2-e^{-6(\alpha \tau)^2} + 3 ~ e^{-2(\alpha \tau)^2}\right)~.
\end{equation}
From Eqs. (\ref{fid_1q_bitflip_error}) and (\ref{fid_bit_flip_proj_meas}),
we find that
\begin{equation}
f^g_{\textrm{ec}} > f^g_{\textrm{err}} ~, \quad \textrm{for~all} \quad \alpha\tau < \infty~, \nonumber
\end{equation}
and therefore projective measurement error correction always improves fidelity,
irrespective of the value of $\alpha \tau$.

In our simulations, we varied $\alpha \tau$ (by holding $\alpha$ fixed and
varying $\tau$), and observed how fidelity changes as the measurement strength
$g\tau$ is varied. Our results, averaged over $10^5$ trajectories to cut down
the statistical errors, are shown in Fig. \ref{fig_bfe_f_vs_setau}. We observe
that for large $g\tau$, the fidelity approaches the result obtained by
performing projective measurement error correction. For small $g\tau$, the
measurement current $I_m$ fluctuates heavily, leading to uncertain feedback
that spoils the fidelity. Increasing $g\tau$ reduces these fluctuations,
and the measurement accumulates sufficient information to improve fidelity.
As a consequence, it is not desirable to perform error correction when $g\tau$
is rather small.

\begin{figure}[H]
\centering
\includegraphics[width=0.5\textwidth]{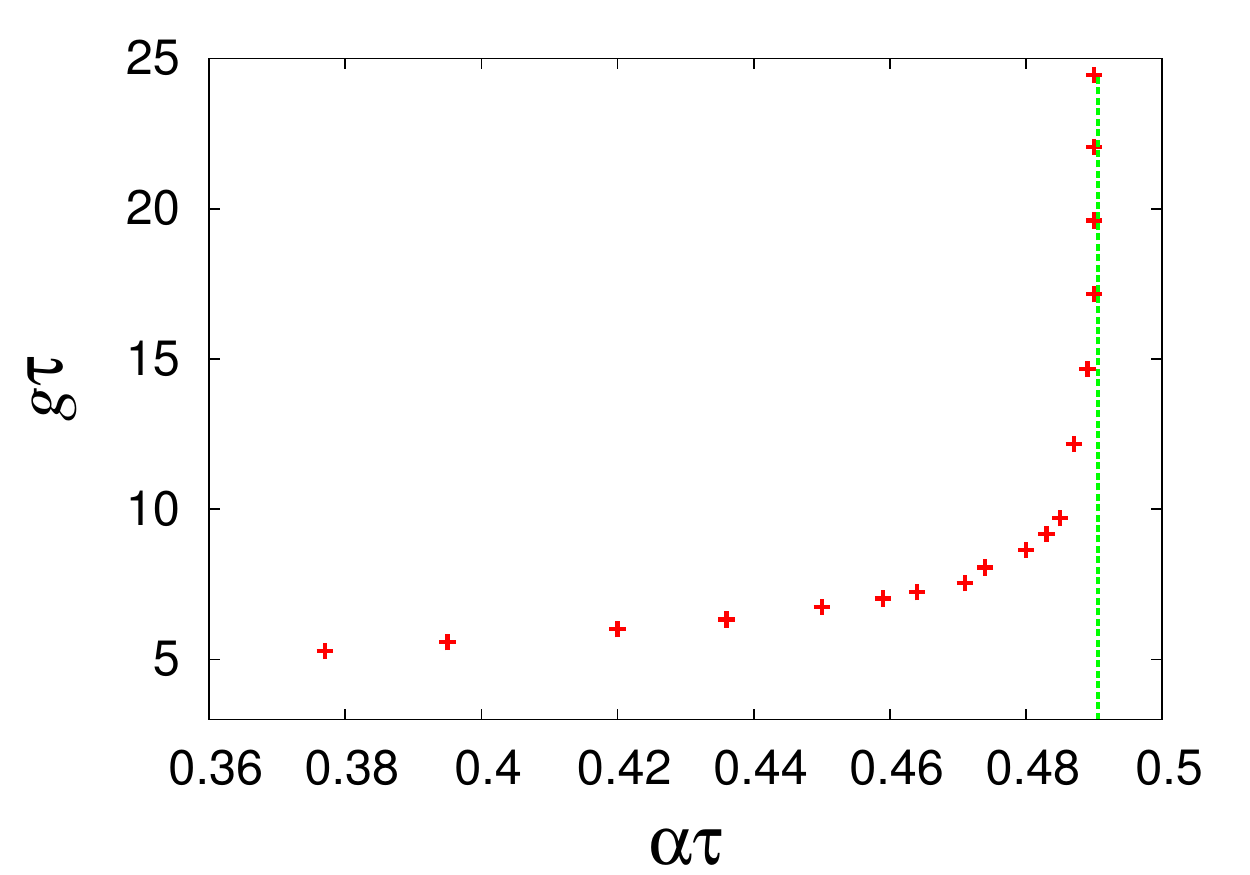}
\caption{Upper bound on the error $\alpha\tau$ for different values of the
measurement coupling, such that $1-f^{(3)}$ reduces by at least a factor of
two after applying error correction. With increasing measurement coupling,
more information about the quantum state can be extracted. That allows the
error correction protocol to cancel larger errors, and the upper bound on
$\alpha\tau$ increases. The green curve shows the upper bound on
$\alpha\tau$ for projective measurement error correction, and it is
approached as $g\tau\rightarrow\infty$.}
\label {ubound_bit_flip_qec}
\end{figure}

\begin{figure}[H]
\centering
\includegraphics[width=0.5\textwidth]{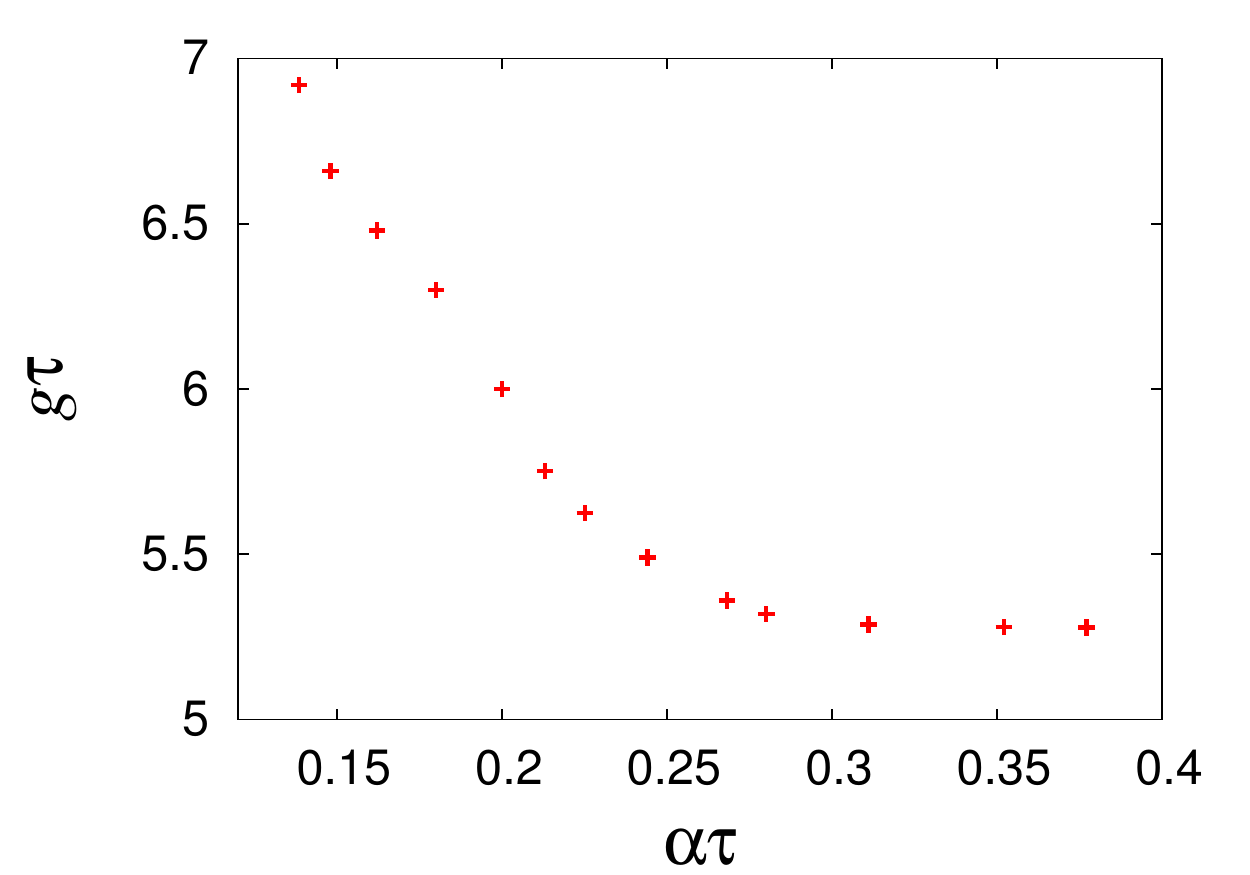}
\caption{Lower bound on the error $\alpha\tau$ for different values of the
measurement coupling, such that $1-f^{(3)}$ reduces by at least a factor of
two after applying error correction. For small measurement couplings, the
quantum state cannot be accurately estimated, unless it is sufficiently
moved away from the logical subspace by a large error. Consequently, the
lower bound on $\alpha\tau$ increases with decreasing $g\tau$.}
\label {lbound_bit_flip_qec}
\end{figure}

It can also be observed from Fig. \ref{fig_bfe_f_vs_setau} that fidelity
improvement varies, depending on the relative strengths of $g$ and $\alpha$.
To illustrate that, we have plotted respectively in Figs.
\ref{ubound_bit_flip_qec}, \ref{lbound_bit_flip_qec}, the upper and lower
bounds on $\alpha\tau$ for different values of $g\tau$, between which
$1-f^{(3)}$ reduces by at least a factor of two after applying error
correction. These two bounds arise for different physical reasons,
and can be understood as follows.

When the measurement coupling decreases, we need to evolve the system for
long $\tau$ to accumulate sufficient information. Within the same duration,
to keep the overall error under control, we need to decrease $\alpha$. As a
result, the upper bound on $\alpha\tau$ decreases with decreasing $g\tau$,
as displayed in Fig. \ref{ubound_bit_flip_qec} (this effect is implicit
in Fig. \ref{fig_bfe_f_vs_setau}). For projective measurement error
correction, the upper bound on $\alpha\tau$, calculated analytically using
Eqs.(\ref{fid_1q_bitflip_error}) and (\ref{fid_bit_flip_proj_meas}), is
0.4905. Fig. \ref{ubound_bit_flip_qec} shows that $\alpha\tau$ approaches
this value as we increase $g\tau$. Beyond this threshold, multiple errors
overwhelm the error correction process and prevent improvement in fidelity.

The lower bound on $\alpha\tau$ results from the error correction protocol
becoming noisy, when inadequate information is extracted by the weak
measurement. In case of projective measurement, the post-measurement state
is known precisely, which allows perfect error correction. But for smaller
$g\tau$, it is difficult to accurately estimate the qubit state from the
measurement current $I_m$ (the approximations described in Section II.B
become far from precise), unless it is sufficiently perturbed by a large
$\alpha\tau$. Consequently the lower bound on $\alpha\tau$ increases
with decreasing $g\tau$, as displayed in Fig. \ref{lbound_bit_flip_qec}
(this effect is also visible in Fig. \ref{fig_bfe_f_vs_setau}).

From our results, we find that the two bounds on $\alpha\tau$ meet around
$g\tau=5.25$. Attempts to correct errors with smaller values of $g\tau$ are
pointless; the best option in such situations is not to perform any error
correction.

\subsection{Single qubit arbitrary error correction}

\begin{figure}
\centering
\includegraphics[width=0.5\textwidth]{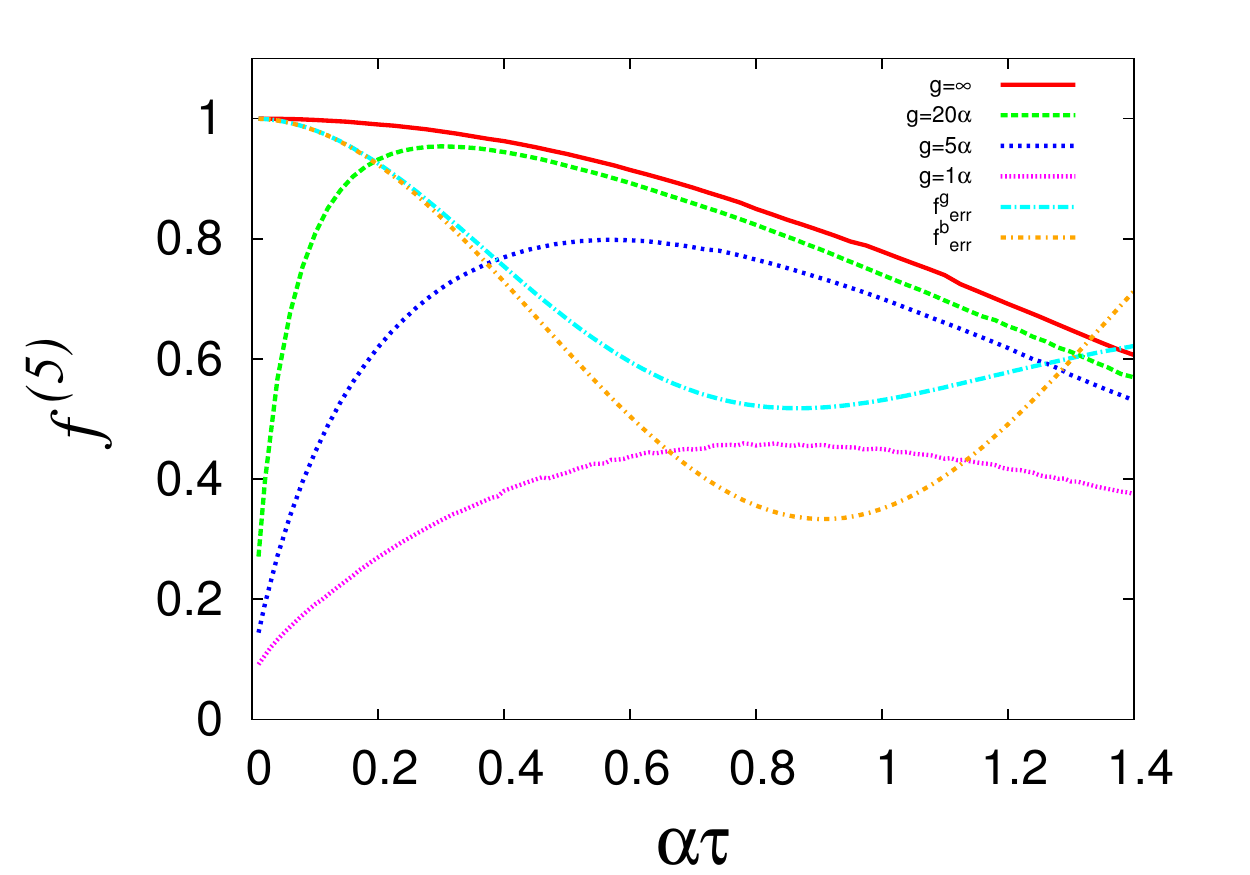}
\caption{Behaviour of the fidelity as a function of the error $\alpha\tau$
for the single qubit arbitrary error correction model. The orange and the cyan
curves show the fidelity in absence of any measurement or error correction,
for binary and Gaussian error distributions respectively. The red curve shows
the fidelity of the five-qubit register, when we perform error correction
using projective measurement. Rest of the curves show fidelity for various
values of the measurement coupling. It is obvious that the fidelity improves
with increasing measurement coupling, and approaches the one obtained by
projective measurement error correction as $g\tau\rightarrow\infty$.}
\label {fid_vs_setau_all_error}
\end{figure}

In the previous subsection, we described how to correct single bit-flip errors.
But a complete quantum error correction protocol has to correct all errors that
may occur. For a single qubit, the complete error Hamiltonian can be written as:
\begin{equation}
\label{ham_1q_all_error}
H_E = \gamma_x \sigma_x + \gamma_y \sigma_y + \gamma_z \sigma_z ~.
\end{equation}
Here $\sigma_x$ represents a bit-flip error, $\sigma_z$ represents a phase-flip
error, and $\sigma_y$ represents both of them occurring together. We take the
error couplings $\gamma_i$ to be independent Gaussian random numbers with zero
mean and variance $\alpha^2$. Assuming that the system is initialized in the
state $|0\rangle$, the density matrix after evolution with the error Hamiltonian
$H_E$ for time $\tau$ becomes,
\begin{widetext}
\begin{eqnarray}
\label{avg_1q_all_error}
\langle\rho(\tau)\rangle &=& \frac{1}{3}
                   \begin{pmatrix}
                   2+(1 - 4 \alpha^2\tau^2) e^{-2(\alpha\tau)^2} & 0 \\
                   0 & 1-(1 - 4 \alpha^2\tau^2) e^{-2(\alpha\tau)^2}
                   \end{pmatrix}~,
\end{eqnarray}
\end{widetext}
when averaged over the distributions of $\gamma_i$. The fidelity of the
quantum state therefore evolves as
\begin{equation}
\label{fid_1q_all_error}
f^g_{\textrm{err}}= \frac{1}{3}\left(2+(1 - 4\alpha^2\tau^2) e^{-2(\alpha\tau)^2}\right) ~.
\end{equation}
If the error distribution is taken to be a binary distribution, instead of a
Gaussian distribution, the averaged density matrix becomes
\begin{eqnarray}
\label{avg_1q_all_error_bm}
\langle\rho(\tau)\rangle &=& \frac{1}{3}
                   \begin{pmatrix}
                   1+2\cos^2 (\sqrt{3}\alpha\tau) & 0 \\
                   0 & 2-2\cos^2 (\sqrt{3}\alpha\tau)
                   \end{pmatrix} ,
\end{eqnarray}
which has the fidelity
\begin{equation}
\label{fid_1q_all_error_bm}
f^b_{\textrm{err}}= \frac{1}{3}\left(1+2\cos^2 (\sqrt{3}\alpha\tau)\right) ~.
\end{equation}
For small values of $\alpha\tau$, Eqs.(\ref{fid_1q_all_error}) and
(\ref{fid_1q_all_error_bm}) give almost the same fidelity as depicted in
Fig. \ref{fid_vs_setau_all_error}, and the choice of error distribution
doesn't matter much.

To protect the single logical qubit from arbitrary error, we encode it in a
five-qubit physical register \cite{LaflammePhysRevLett.77.198,BennettPhysRevA.54.3824},
according to \cite{Preskill}:
\begin{eqnarray}
\label{5qenc}
|0\rangle_L &\rightarrow& |00000\rangle_P + |10010\rangle_P + |01001\rangle_P + |10100\rangle_P \nonumber\\ &+& |01010\rangle_P + |00101\rangle_P - |11110\rangle_P - |01111\rangle_P \nonumber\\ &-& |10111\rangle_P - |11011\rangle_P - |11101\rangle_P - |01100\rangle_P \nonumber\\ &-& |00110\rangle_P - |00011\rangle_P - |10001\rangle_P - |11000\rangle_P , \nonumber\\
\label{5qenc1}
|1\rangle_L &=& \bar{X}|0\rangle_L ,
\end{eqnarray} 
where $\bar{X} = XXXXX$. In our simulations, without loss of generality,
we choose the initial state to be $|00000\rangle$, which makes our measured
fidelity of the encoded state
\begin{equation}
\label{fid_enc_1q_arbitrary_flip}
f^{(5)}={}_P\langle 00000|\rho|00000\rangle_P ~.
\end{equation} 
We diagnose all single Pauli errors using the four syndrome operators:
\begin{eqnarray}
\label{5qmo}
M_1 &=& XZZXI, \nonumber \\
M_2 &=& IXZZX, \nonumber \\
M_3 &=& XIXZZ, \nonumber \\
M_4 &=& ZXIXZ. 
\end{eqnarray}

For the five-qubit register, the independent error Hamiltonian can be written
as:
\begin{eqnarray}
\label{5qeh}
H_E &=& \gamma_1 (XIIII) + \gamma_2 (IXIII) + \gamma_3 (IIXII) \nonumber \\
 &+&  \gamma_4 (IIIXI) + \gamma_5 (IIIIX) + \gamma_6 (YIIII) \nonumber \\
 &+&  \gamma_7 (IYIII) + \gamma_8 (IIYII) + \gamma_9 (IIIYI) \nonumber \\
 &+&  \gamma_{10} (IIIIY) + \gamma_{11} (ZIIII) + \gamma_{12} (IZIII) \nonumber \\
 &+&  \gamma_{13} (IIZII) + \gamma_{14} (IIIZI) + \gamma_{15} (IIIIZ),
\end{eqnarray} 
where the error couplings $\gamma_i$ are independent Gaussian random numbers
with zero mean and variance $\alpha^2$.

Our feedback Hamiltonian is: 
\begin{eqnarray}
\label{5qfh}
H_F &=& \lambda_1 (XIIII) + \lambda_2 (IXIII) + \lambda_3 (IIXII) \nonumber \\
 &+&  \lambda_4 (IIIXI) + \lambda_5 (IIIIX) + \lambda_6 (YIIII) \nonumber \\
 &+&  \lambda_7 (IYIII) + \lambda_8 (IIYII) + \lambda_9 (IIIYI) \nonumber \\
 &+&  \lambda_{10} (IIIIY) + \lambda_{11} (ZIIII) + \lambda_{12} (IZIII) \nonumber \\
 &+&  \lambda_{13} (IIZII) + \lambda_{14} (IIIZI) + \lambda_{15} (IIIIZ),
\end{eqnarray}
where $\lambda_i$ are the feedback couplings. With four different binary
syndrome measurements, there are 16 possible outcomes. The one with all four
currents positive stands for ``no error'', while the other 15 possibilities
correspond to single qubit $X,Y,Z$ errors. Different combinations of the
measured currents determine the corresponding non-zero $\lambda_i$, as listed
in the following table.
\begin{center}
\begin{tabular} {|c|c|c|c|c|}
\hline

$I_m^1$ & $I_m^2$ & $I_m^3$ & $I_m^4$ & Non-zero $\lambda_i$ \\
 \hline
+ & + & + & - & $\lambda_1$ \\
\hline
- & + & + & + & $\lambda_2$ \\
\hline
- & - & + & + & $\lambda_3$ \\
\hline
+ & - & - & + & $\lambda_4$ \\
\hline
+ & + & - & - & $\lambda_5$ \\
\hline
- & + & - & - & $\lambda_6$ \\
\hline
- & - & + & - & $\lambda_7$ \\
\hline
- & - & - & + & $\lambda_8$ \\
\hline
- & - & - & - & $\lambda_9$ \\
\hline
+ & - & - & - & $\lambda_{10}$ \\
\hline
- & + & - & + & $\lambda_{11}$ \\
\hline
+ & - & + & - & $\lambda_{12}$ \\
\hline
+ & + & - & + & $\lambda_{13}$ \\
\hline
- & + & + & - & $\lambda_{14}$ \\
\hline
+ & - & + & + & $\lambda_{15}$ \\
\hline
\end{tabular}
\end{center}

We estimate the feedback rotation angle by extending the procedure described
in Section II.C to four binary measurements. At every evolution step, only
one (or none) of the fifteen feedback couplings is non-zero, determined by
its unique syndrome signature. The non-zero rotation angle always equals
$-\frac{\bar{\theta}}{2\tau}$. 

In our simulations, we once again varied $\alpha \tau$ (by holding $\alpha$
fixed and varying $\tau$), and observed how the fidelity changes as the
measurement strength $g\tau$ varies. Our results, averaged over $10^5$
trajectories to control the statistical errors, are shown in Fig.
\ref{fid_vs_setau_all_error}. We notice the same overall features as in the
case of the bit-flip error. For large $g\tau$, as expected, the fidelity
approaches the value for projective measurement error correction. For small
$g\tau$, large fluctuations of the measurement current $I_m$ make the feedback
uncertain and spoil the fidelity. With increasing $g\tau$, these fluctuations
reduce and the fidelity improves, while error correction with rather small
$g\tau$ is to be avoided.

\begin{figure}
\centering
\includegraphics[width=0.5\textwidth]{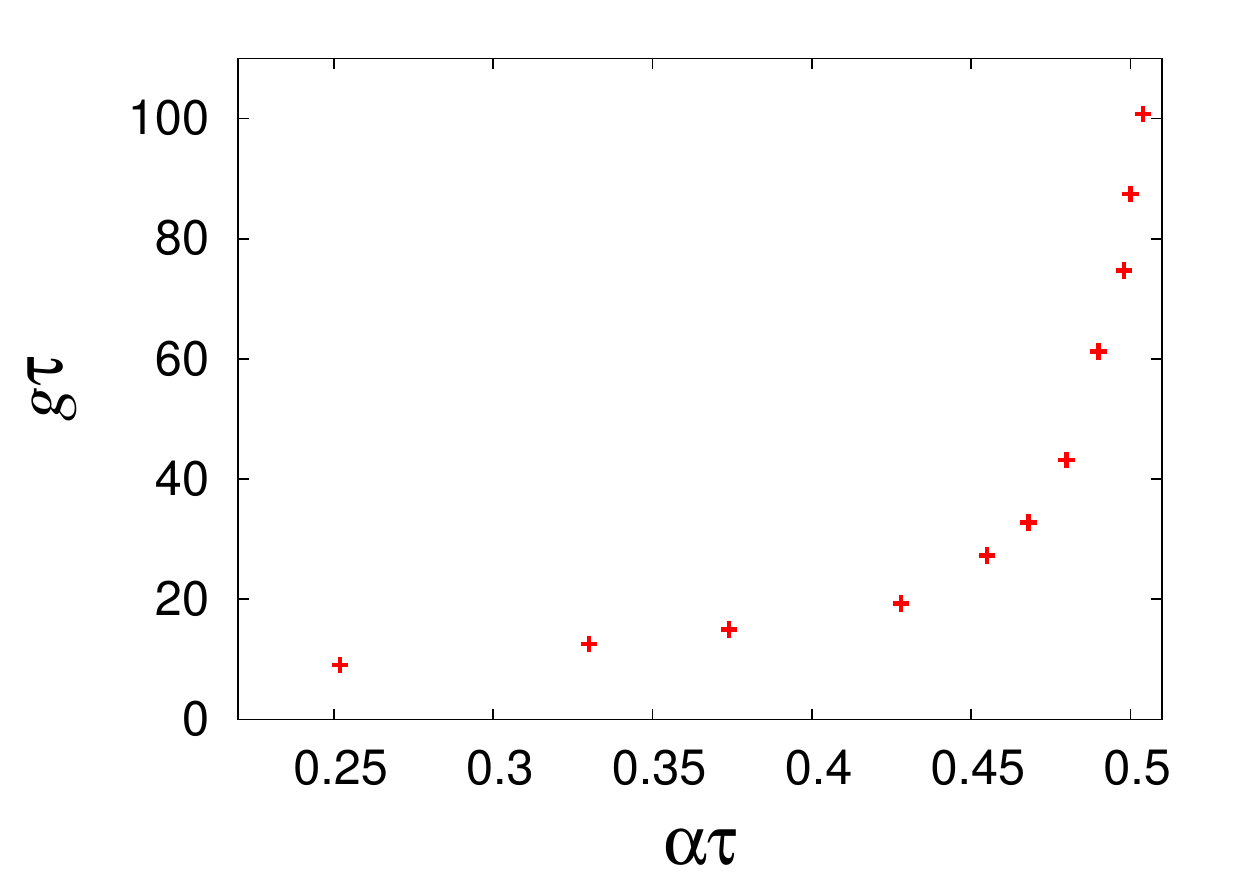}
\caption{Upper bound on the error $\alpha\tau$ for different values of the
measurement coupling, such that $1-f^{(5)}$ reduces by at least a factor of
two after applying error correction. With increasing measurement coupling,
more information about the quantum state can be extracted. That allows the
error correction protocol to cancel larger errors, and the upper bound on
$\alpha\tau$ increases. The upper bound for projective measurement error
correction is $\alpha\tau=1.025$, and is outside the range of this figure.}
\label {ubound_all_error_qec}
\end{figure}

\begin{figure}

\includegraphics[width=0.5\textwidth]{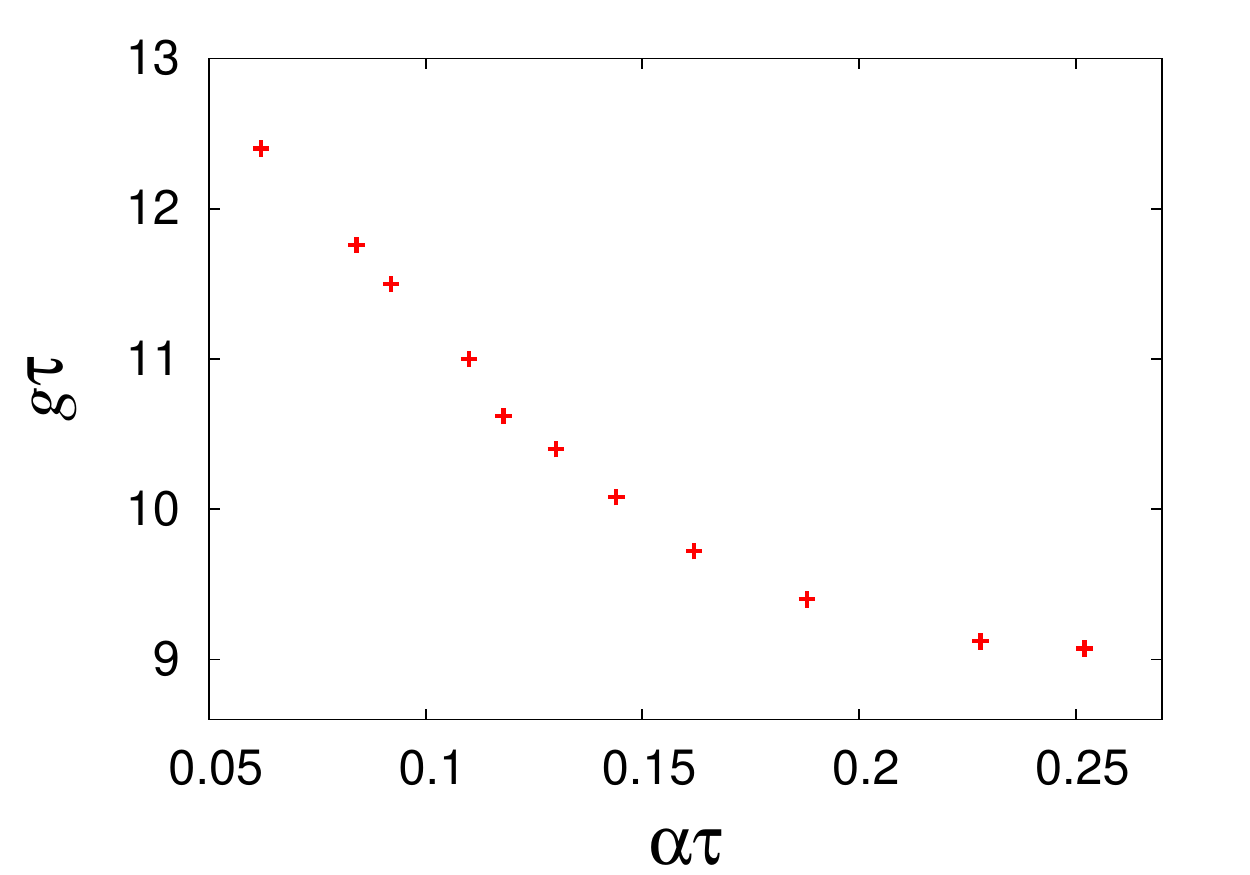}
\caption{Lower bound on the error $\alpha\tau$ for different values of the
measurement coupling, such that $1-f^{(5)}$ reduces by at least a factor of
two after applying error correction. For small measurement couplings, the
quantum state cannot be accurately estimated, unless it is sufficiently
moved away from the logical subspace by a large error. Consequently, the
lower bound on $\alpha\tau$ increases with decreasing $g\tau$.}
\label {lbound_all_error_qec}
\end{figure}

Improvement of the fidelity varies depending on the relative strength of
$g$ and $\alpha$, which can be observed in Fig.\ref{fid_vs_setau_all_error}.
To make that more explicit, we have plotted respectively in Figs.
\ref{ubound_all_error_qec} and \ref{lbound_all_error_qec}, the upper and
lower bounds on $\alpha\tau$ for different values of $g\tau$, between which
$1-f^{(5)}$ reduces by at least a factor of two after applying error correction.
The physical reasons for these bounds are the same as those described for the
bit-flip error correction scheme in the previous subsection. The upper bound
specifies the threshold beyond which error correction fails due to multiple
errors, and the lower bound signifies the minimum information to be extracted
by measurement in order to perform error correction. The two bounds meet
around $g\tau=9.1$, and it is not worthwhile to attempt error correction for
smaller values of $g\tau$.

In case of projective measurement, we numerically find that error correction
cannot improve fidelity beyond $\alpha\tau = 1.375$. When we demand a
factor of two improvement in $1-f^{(5)}$, this threshold decreases to
$\alpha\tau = 1.025$. Our results approach this upper bound rather slowly
as $g\tau$ increases. This behaviour sharply contrasts with the fast approach
to the upper bound in case of the bit-flip error correction protocol. It may
be that the considerably larger error subspace of the five-qubit register
requires the measurement to extract more information in order to cut down
the error.

\section{Discussion}

We have constructed an error correction protocol based on results of weak
measurements of qubits, and numerically tested it as a function of the
measurement coupling. Our error model consists of random Gaussian fluctuations,
which are common in real life situations. We have shown that it is possible
to improve the fidelity of the encoded logical state with our protocol,
provided that the error rate is below the threshold beyond which multiple
errors dominate, and the measurement strength is large enough to extract
sufficient information to perform error correction. We have expressed these
features as upper and lower bounds on the error size $\alpha\tau$, for
various values of the measurement strength $g\tau$, between which error
correction succeeds. This range of $\alpha\tau$ is maximized for projective
error correction, i.e. $g\tau\rightarrow\infty$. So projective error
correction is always preferable, whenever it is possible. In case projective
error correction is not possible, in physical systems where measurement would
take sizeable time, we can use weak measurements to improve fidelity of the
quantum state. Even then, the effort is fruitful only when $g\tau$ exceeds
certain minimum value. Our simulations have obtained this minimum value for
single qubit bit-flip and arbitrary error correction codes. Error correction
with smaller values of $g\tau$ should be avoided.

On quite general grounds, we can express the combined state of the system
and the ancilla in a form that separates the logical and the error subspace
components:
\begin{equation}
|\psi_L\rangle|a_L\rangle + |\psi_E\rangle_\perp|a_E\rangle ~.
\nonumber
\end{equation} 
Here $|\psi_L\rangle$ and $|\psi_E\rangle$ are normalised system states in
the logical and error subspaces respectively, while $|a_L\rangle$ and
$|a_E\rangle$ are some unnormalised states of the ancilla. The magnitude of
$|a_E\rangle$ determines the fidelity of the encoded quantum state. Initially,
this magnitude is zero, but it becomes $O(\gamma)$ after evolution under the
error Hamiltonian. After applying the error correction feedback but before
resetting the ancilla, the joint state of the system and the ancilla is
entangled. At this stage, the magnitude of $|a_E\rangle$ is $O(\gamma^2)$
for projective measurement error correction, but it remains $O(\gamma)$ for
weak measurement error correction due to only partial elimination of the
error. Ultimately, resetting the ancilla makes the encoded quantum state
mixed, and the magnitude of $|a_E\rangle$ is a measure of the deviation
from purity. Our analysis has shown that although weak measurements cannot
produce as good quantum error correction as projective measurements, they
do manage to reduce the magnitude of $|a_E\rangle$ under certain conditions
and so can be useful.

Performing error correction using weak measurements is becoming feasible,
with technological advances in superconducting transmon systems
\cite{reed2012realization}. It would be interesting to check our proposal
in such experiments.

\section*{Acknowledgments}
PK is supported by a CSIR research fellowship from the Government of India.
We are grateful to Rajamani Vijayaraghavan for useful discussions and helpful
comments on the earlier draft of this work.

\bibliography{refQECweakmeas}

\end{document}